# Modelling and Using Response Times in Online Courses

**Ilia Rushkin**
Harvard University
Cambridge, MA, USA
ilia_rushkin@harvard.edu

**Isaac Chuang**
Massachusetts Institute of Technology
Cambridge, MA, USA
ichuang@mit.edu

**Dustin Tingley**
Harvard University
Cambridge, MA, USA
dtingley@gov.harvard.edu

**ABSTRACT**: Each time a learner in a self-paced online course seeks to answer an assessment question, it takes some time for the student to read the question and arrive at an answer to submit. If multiple attempts are allowed, and the first answer is incorrect, it takes some time to provide a second answer. Here we study the distribution of such "response times." We find that the log-normal statistical model for such times, previously suggested in the literature, holds for online courses. Users who, according to this model, tend to take longer on submits are more likely to complete the course, have a higher level of engagement, and achieve a higher grade. This finding can be the basis for designing interventions in online courses, such as MOOCs, which would encourage "fast" users to slow down.

**NOTES FOR PRACTICE** (for Research Papers Only)

- Response time is a type of data that complements response correctness (or score) in learning analytics. Similar to how response correctness data is transformed into learner's latent ability or skill mastery, response time can be transformed into learner's slowness (or, alternatively, speed). Recently, several learner models have been proposed that make use of response times.
- We find a way to extract user response times on assessment questions in HarvardX courses and fit them using a log-normal model via likelihood maximization. The model quantifies each user's slowness on assessment questions, controlling for each question's intrinsic time-intensity.
- We find that a user's tendency to be slower on assessment items is correlated with higher completion rates, higher final grades in the course, and higher certification rates. This provides a basis for future interventions in online courses that would encourage fast users to slow down.
- Our model also outputs characteristics of each assessment item, which are of interest to course content creators.

## 1. INTRODUCTION

When users interact with assessment questions in an online course, the data that usually receives the most attention is the answers they submit, and sometimes only the correctness of those answers or their received score. But the time the user spends on the question is also important: it is arguably the most readily acquired data that reveals something about the process by which the user arrived at an answer. Analyzing these "response times" allows one to quantify some properties of the questions (how long does a question typically take and how much does it vary?) as well as some properties of the users (how much time do they tend to take in answering questions?). The question properties have implications for course design, and user speed may be related to the user's ability and preferred mode of interaction with the course. Extracting such parameters necessitates a parametric statistical model for the response times. This is similar to how in IRT (item response theory) an item response function is needed for extracting question parameters and user ability from the response correctness data (Hambleton, Swaminathan, & Rogers, 1991; Baker & Kim, 2004).

In this paper, we study "user slowness," an interesting and little-used parameter that we extract from user response times as a measure of how long it takes a user to respond to a question. It can be interpreted in two fundamentally different ways: 1) taking a longer time could be viewed as a sign of lower user mastery, and 2) it could be viewed as a sign of diligence and thoroughness. In the first interpretation, longer response times should be associated with lower achievement; in the second, the opposite is expected. Which relationship applies in reality is not *a priori* obvious, most likely depending on the context. If users solve problems slowly in a timed test, the first interpretation could be more likely, since they are under pressure to give answers as quickly as they can work them out with reasonable certainty. But in a self-paced[1] course, such as a MOOC, the second could be more applicable. Our findings below demonstrate a correlation between user slowness and success in the course. It implies that, if course instructors track learner slowness in responses, the fast-responding learners demonstrate a greater cause for intervention than the slow ones. Moreover, the interpretation of slowness as *thoroughness* suggests causality rather than just correlation. If so, user slowness in a self-paced online course is a desirable quality. Furthermore, it is a feature that can potentially be manipulated by course instructors who could issue a recommendation to slow down to a student who is going fast and not performing well.

User slowness (or any other way of including the response time data) is a valuable extra dimension in learner modelling. It complements the information about response correctness, normally used to estimate a learner's state of knowledge. It can, for instance, expose the distinction between the mastery of a skill and the fluency of the skill application (Wang, Zhang, Douglas, & Culpepper, 2018). However, little information is available about using response times in a self-paced course environment. We might expect that the pattern of response times should be directly affected by the lack of time pressure and the variety of content (not just questions or tasks), as compared to a timed test, or a test in which the learner works through a task sequence of escalating complexity. Our study aims to fill both these gaps.

---

[1] By "self-paced," we mean that the users who submit answers to questions are not subject to the time pressure of a timed test, in which students are supposed to perform a certain number of tasks within a set time. None of the HarvardX courses used in this study employed such timed tests, although weekly homework deadlines existed in some courses.

Furthermore, there is a problem of estimating response times from the event stream. If questions are served one by one, it is a trivial matter of taking the difference between timestamps of question loadings and submit events. But in online courses, multiple questions may be served on the same page, so that the extraction of raw response times from the event stream data becomes a significant methodological step. Below we suggest a simple practical way of estimating response times in such cases.

This paper is organized as follows: Section 2 introduces the model and Section 3 situates it among other related works. Section 4 describes the data to which we applied the model. Section 5 discusses how well the model describes the data, as well as some insights on the outputs of the model and how they depend on the correctness of the submitted answer or the attempt number. In Section 6, we demonstrate that user slowness, extracted by the model, can be used as a predictor of the user's success in the course. In Section 7, we look into the relation of user slowness with user demographics and several other variables describing their activity in the course.

## 2. MODEL DESCRIPTION

We assume that the response time of any user on any assessment question is a random variable such that its logarithm has normal distribution. This is known as a log-normally distributed variable. On a basic level, we choose the log-normal distribution for the same reasons one might choose the normal distribution to model any histogram with a relatively un-skewed bell shape. The log-normal distribution is a model of some convenience, as it is familiar, easy to work with, and has qualitatively correct features: no negative values in the domain, a single peak, and a long tail on the right. But there is a deeper reason for log-normality. The central limit theorem (covered in most introductions to probability and statistics, e.g., Grimmett & Stirzaker, 2001) states that, under some mild conditions, the sum of a very large number of independent random variables approaches normal distribution. This is the reason for the ubiquity of normal distributions in nature, because observed quantities are often the sum of many independent random contributions. Should such contributions be multiplicative, rather than additive, they would give rise to a log-normal distribution, and this seems a reasonable idea when dealing with response times.

To see the multiplicative nature of the process of responding to assessment questions, suppose there is a certain basic response time $t_B$ for a user-question interaction (longer for harder questions and for slower users). The actual response time is affected by a large number of diverse factors, such as having to think about different aspects of the question, calculations, looking up information, fatigue, distractions, etc. The extra time taken up by any factor should scale with the difficulty of the question and with the overall slowness tendency of the user, i.e., with the basic time $t_B$. Therefore, it is natural to assume that the effect of each factor is multiplicative: a factor $i$ multiplies the basic response time by $(1 + r_i)$, where $r_i$ is a random variable ("rate"), resulting in the response time $t = t_B \prod_i (1 + r_i)$. In this setting, the central limit theorem predicts that the distribution of $t$ will approach the log-normal distribution when the number of contributing factors is large.[2]

Following van der Linden (2006) and Bertling and Chuang (2015), we model the response time logarithms as independent normally distributed variables with probability density:

$$P(\ln t_{qu}) = \frac{\alpha_q}{\sqrt{2\pi}} \exp\left(-\frac{\alpha_q^2}{2}\left(\beta_q + \zeta_u - \ln t_{qu}\right)^2\right), \tag{1}$$

where $q = 1, 2, \ldots N_q$ is a question, $u = 1, 2, \ldots N_u$ is a user, and $t_{qu}$ is the response time of user $u$ on question $q$. The user parameter $\zeta_u$ is the "user slowness." The question parameters $\alpha_q$ and $\beta_q$ are interpreted as "discrimination" and "time intensity." The question discrimination $\alpha_q$ measures the size of random fluctuations of the response times around the expectation value: high discrimination means small fluctuations and vice versa. In other words, it is the measure of the question's sensitivity to the variability in learner speed. "Time intensity" is a type of difficulty measure for each question. Conceptually, this model is somewhat analogous to item response theory (Hambleton et al., 1991; Baker & Kim, 2004). There too, the user-question interaction is modelled by combining a set of user parameters (latent ability) and a set of question parameters (discrimination, difficulty, and possibly guess and slip probabilities).

In Eq. 1, there is a freedom of shifting all $\beta$'s and all $\zeta$'s by opposite constants without affecting the probability distribution. We fix this freedom by imposing the condition $\sum_u \zeta_u = 0$. Thus, if the response times are measured in seconds, $\exp(\beta_q)$ is question $q$'s characteristic response time in seconds and $\exp(\zeta_u)$ is the multiplicative factor by which user $u$'s response times tend to differ from those characteristic response times.[3]

In this way, the model is defined by $2N_q + N_u - 1$ free parameters, and the number of observed values $t_{qu}$ scales as $N_u \times N_q$ (more or less, since not all users respond to all questions). For substantial numbers of users and questions, the number of observations will be much greater than the number of parameters, making it possible to fit the parameters by maximizing likelihood. Namely, given the observed response times $t_{qu}$ in the set of observations $(q, u) \in \mathcal{O}$ we find the parameters of the questions and the slowness of the users via minimization of negative logarithmic likelihood:[4]

---

[2] We assume that the conditions of the theorem are fulfilled. In practice, the most vulnerable condition of the theorem is that the variables $x_i = \ln(1 + r_i)$ should be independent, or at least not universally non-independent (they could form distinct independent groups with high internal correlation, but then the number of such groups needs to be large), which is the mathematical expression of the assumption that the nature of the variables is diverse.

[3] The mean time intensity across questions equals the mean expected logarithm of response times across all questions and users: $N_q^{-1} \sum_q \beta_q = N_q^{-1} N_u^{-1} \sum_{q,u} (\beta_q + \zeta_u)$.

[4] The code we used for this on HarvardX data is open source and available at https://github.com/harvard-vpal/log-normal-response-time

$$\{\alpha, \beta, \zeta\} = \operatorname{argmin} \sum_{(q,u)\in O} \left( \frac{\alpha_q^2}{2} \left(\beta_q + \zeta_u - \ln t_{qu}\right)^2 - \ln \alpha_q \right). \tag{2}$$

The minimization needs to be restricted to positive values $\alpha_q$, whereas $\beta_q$ and $\zeta_u$ are unconstrained.

## 3. RELATED WORK

The use of logarithms of response times (rather than response times themselves) and of fitting the time data with a log-normal distribution is at least as old as 1983. In Thissen's (1983) study, response time logarithms are combined with the parameters of the item response theory to model the trade-off between speed and accuracy as well as the relation between the time intensity of a question and its difficulty. The idea of incorporating response times (logarithms or not) into the framework of item response theory has been investigated.[5] Inclusion of response times enriches the IRT model by tracking the fluency of skill application in addition to skill mastery (Wang et al., 2018). Further extra variables are also possible: Beck (2005) combined item response theory with response time and with question length to model student disengagement. On the other hand, there exist studies (Schnipke & Scrams, 1999; van der Linden, Scrams, & Schnipke, 1999) that model response times without incorporating any response variables such as IRT parameters. The specific form of the model used here as Eq. 1 was first investigated by van der Linden (2006).

It should be noted that time distributions other than log-normal have also been tried. Notably, in Scheiblechner (1979) and Scheiblechner (1985), the distribution is exponential: $P(t_{qu}) = \lambda_{qu} \exp(-\lambda_{qu} t_{qu})$, where the distribution parameter is taken to be a sum of a question-specific parameter and a user-specific parameter $\lambda_{qu} = \theta_u + \epsilon_q$. The implied assumption is that the problem-solving process is modelled as waiting for an epiphany, which is equally likely to occur at any time (probability $\lambda_{qu} dt$ for any infinitesimal time interval $dt$), and the problem is submitted as soon as it happens. This is appropriate for some types of mental activity (e.g., recalling facts or solving riddles), but clearly not for submitting questions in an online course: here, the observed distributions of response times invariably have a shape that qualitatively resembles a log-normal distribution rather than an exponential one. Maris (1993) proposed a generalization to gamma distributions, which are a family of statistical distributions, whose probability density is a product of an exponential and a power-law: $\propto t^{\alpha-1} \exp(-\beta t)$, where both parameters $\alpha, \beta$ are assumed positive. The exponential distribution is a special case of gamma distribution with $\alpha = 1$. For $0 < \alpha \leq 1$ the shape of the gamma distribution resembles that of the exponential one, but for higher values of $\alpha$ the behavior changes and begins to resemble a log-normal one: the probability density increases from $t = 0$ to form a single peak, after which it decays in a long tail. In particular, for $\alpha > 2$ both the density and its derivative vanish at $t = 0$, as they do in the log-normal distribution.

---

[5] For an overview of adding response times to the Rasch model (a variant of item response theory) and an implementation of the speed-accuracy trade-off function and conditional accuracy functions in the context of timed tests, see Roskam (1987), Roskam (1997), and Verhelst, Verstralen, and Jansen (1997).

A simple "binning" approach to response times is also possible, without any assumptions about the shape of their distribution. Lin, Shen, and Chi (2016) describe the use of Bayesian knowledge tracing (BKT), to which a binary variable ("quick/slow") is added to describe the response time: for each item, the median of response times is calculated, and responses are labelled quick (slow) if they are below (above) the median. This is a simple way of leveraging some information about learner speed for estimating their mastery.

Originally, the log-normal model of response times was developed for test items, and it continues to be applied in this setting. Recently, Zhan, Jiao, and Liao (2018) combined it with the DINA model for response correctness in application to the PISA 2012 data (computer-based high-school mathematics test). Wang et al. (2018) also studied the application to the spatial rotation test, in particular tracing how the slowness (latent speed) changes as the learner goes through the test. As anticipated, the learner's latent speed tends to grow, and can serve as an indicator of developing fluency. To our knowledge, the model was first applied to assessment items in MOOCs in the unpublished work of Bertling and Chuang (2015), where the main direction of the investigation is in linking user slowness and IRT latent ability.

## 4. METHODS

We use a non-linear conjugate-gradient routine (Dai & Yuan, 2001), implemented in the R package "Rcgmin," to perform the minimization from Eq. 2 for 47 HarvardX courses from 2015–2017, which involve more than 34,000 learners and 4,000 assessment questions in total. Among these, there were 16 STEM courses (natural and health sciences, computing, and programming) and 31 non-STEM courses (humanities, law, social sciences). To reduce the number of responses from non-committed users, we restricted the data to those users who visited at least half of the chapters in the course. It is a standard measure in HarvardX data analysis, where such users are said to have "explored" the course (Ho et al., 2014). Further, we discard instances when a user submitted more than five answer attempts, as a large or even unlimited number of attempts might provoke a different, guess-driven behavior. We also discard questions answered by fewer than 10 users, which provides an insufficient amount of data. Similarly, we want to avoid users who respond to only a few questions; to this end, we impose a 10-question minimum here as well.[6] We can call the questions and users who remain in the data after this procedure "qualified."

Because questions in HarvardX courses often allow multiple submit attempts, we attempted to fit the model in each course on first and second submits separately. In doing so, we take care to include second submits only if the response on the first submit was incorrect (second responses after correct first responses are understandably rare, but they do occur, so we remove them from the data and keep only the user's first response for that question). It should be remembered that the data on second responses entails selection bias: not all questions allowed multiple submits; when they did, the submits occurred only after the incorrect first response.

---

[6] In a few courses, we found fewer than 10 questions in total, for any user, so we lowered the cut-off to the maximum encountered value. In such a course, we assume that an engaged user should submit all of the available questions, and in fact, a large proportion of users do. Simply removing all such courses from our dataset does not affect our findings.

Our model deals only with the times of responses, not with their correctness. Whether the response was right or wrong should not significantly affect the outcome of the model. In order to check that, we fitted our model separately to three groups of submit events: 1) all submits regardless of their correctness, 2) only correct responses, and 3) only incorrect responses. The first group is the intended way of applying the model; the other two are for comparison only. Repeating this process for first and second submits, we produce up to six model fits for each course.

Because our interest here is in response time, the decision about how to measure this variable is critical. In principle, this is the difference between 1) the time the question appears on the screen and 2) the timestamp of the first submit click. The time the question appears can be determined by the timestamp of the user loading the question page. The challenge is that sometimes multiple questions are served on the same page, and because a typical user works through them in sequence, the page-loading timestamp of these questions will artificially lengthen the first response time for all the questions on the page except the first one responded to. Our strategy to resolve this problem can be described as follows: in case of multiple questions on a page, assume that the user starts working on a question after the chronologically last submit click for a different question from the same page. Namely, suppose we observe in the data that, for a given user, a group of questions has the same page-loading timestamp $t_0$, and the submit timestamps are arranged and indexed chronologically as $t_0 < t_1 < t_2 < \cdots < t_p$. These submit events belong to different questions, possibly with multiple submits on a question, and it is not assumed that the user works on questions completely sequentially (e.g., it can be that $t_1$ is the first submit on question A, $t_2$ is the first submit on question B, $t_3$ is the second submit on question A again). If the timestamp of the first submit for one of the questions is $t_i$ ($i > 0$), then the first response time for this question is calculated as $t_i - t_{i-1}$.[7]

Our definition of the second response time is simply the difference between the timestamps of the first and the second submit clicks, which assumes that the user starts thinking about the second answer immediately after seeing that that first one was incorrect. The learning platform does not insert a pause or an intermediate step (such as, providing a hint or modifying the question) before the user can make another submit.

After preparation, the data for each of the 6 model fits (correct/incorrect/any responses on first/second attempts) in a course is in the form of an $N_u \times N_q$ matrix, where each row is a qualified user, each column is a qualified question, and the entries are the natural logarithms of times in seconds (except where data is missing). Since the measurement of second response times uses fewer assumptions, it may seem more reliable. However, second responses occur only when the question allows more than one attempt and the first response was incorrect (which in the case of a partially correct answer involves an extra dichotomizing step), meaning a smaller and possibly biased data sample. For these reasons, we regard first response times of any correctness as the most valuable subset of data. Its data matrix is guaranteed to have the biggest dimensions and the most data. Other matrices contain fewer observations. Convergence on the data from first responses of any correctness was achieved in 45 out of 47 courses, but only in 21 courses on the data from second responses of any correctness. When aggregating the data across courses, we include only the converged fits. Table 1 lists some parameters related to the amount of data available.

---

[7] We do not impose any timeout cut-off on the response times, as the model is supposed to make sense of any time-values without supervision. Only about 7% of first response times and 0.9% of second response times in our entire dataset exceed 24 hours. The median response times in the dataset are 112 seconds for first responses and 17 seconds for second responses.

**Table 1: Dataset Parameters for First and Second Responses of Any Correctness across Courses**

| Question response | Statistic | min | median | max | total |
|---|---|---|---|---|---|
| First | Number of users ($N_u$) | 13 | 567 | 3,055 | 34,105 |
| | Number of questions ($N_q$) | 7 | 61 | 447 | 4,020 |
| | Missingness ($m$) | 0.03 | 0.25 | 0.72 | N/A |
| | $r$ | 0.009 | 0.035 | 0.171 | N/A |
| Second | Number of users ($N_u$) | 13 | 674 | 7,628 | 27,118 |
| | Number of questions ($N_q$) | 17 | 63 | 316 | 2,107 |
| | Missingness ($m$) | 0.62 | 0.76 | 0.89 | N/A |
| | $r$ | 0.035 | 0.085 | 0.674 | N/A |

**Note:** $\boldsymbol{N_u}$ is the number of users (rows) in the data, $\boldsymbol{N_q}$ is the number of questions (columns), $\boldsymbol{m}$ is missingness (the fraction of missing matrix entries), and $\boldsymbol{r} = (\boldsymbol{2N_q} + \boldsymbol{N_u} - \boldsymbol{1})/(\boldsymbol{N_u N_q}(\boldsymbol{1} - \boldsymbol{m}))$ is the ratio of the number of fit parameters to the number of observations. The last column provides the total numbers of learners and questions from all courses. Only the cases where convergence was reached are included.

## 5. ASSESSING MODEL QUALITY

After the fit, we check how closely our model approximates the response times and the extent of the influence of response correctness on the model outputs.

The model assumes that the variables $x_{qu} = \alpha_q(\ln t_{qu} - \beta_q - \zeta_u)$ should be standard normal variables, and so we can plot the observed cumulative distribution (percentile curve) $\mathrm{CDF}(x)$ vs. the cumulative distribution of the standard normal variable $\Phi(x)$. The result is shown in Figure 1, where we list the first four moments with respect to the origin: $m_k = (1/|\mathcal{O}|)\sum_{(q,u)\in\mathcal{O}}(x_{qu})^k$ (the standard normal distribution has $m_1 = 0$, $m_2 = 1$, $m_3 = 0$, $m_4 = 3$). Although the deviations are noticeable, it should be kept in mind how large the range of times is: the interquartile range is from 30 to 670 seconds for first response times and from 6 to 44 seconds for second response times. It is clear that time-logarithms are suitable variables for analysis: the skewness of distribution of response times themselves is extreme, but logarithmic transformation accounts for virtually all of it.

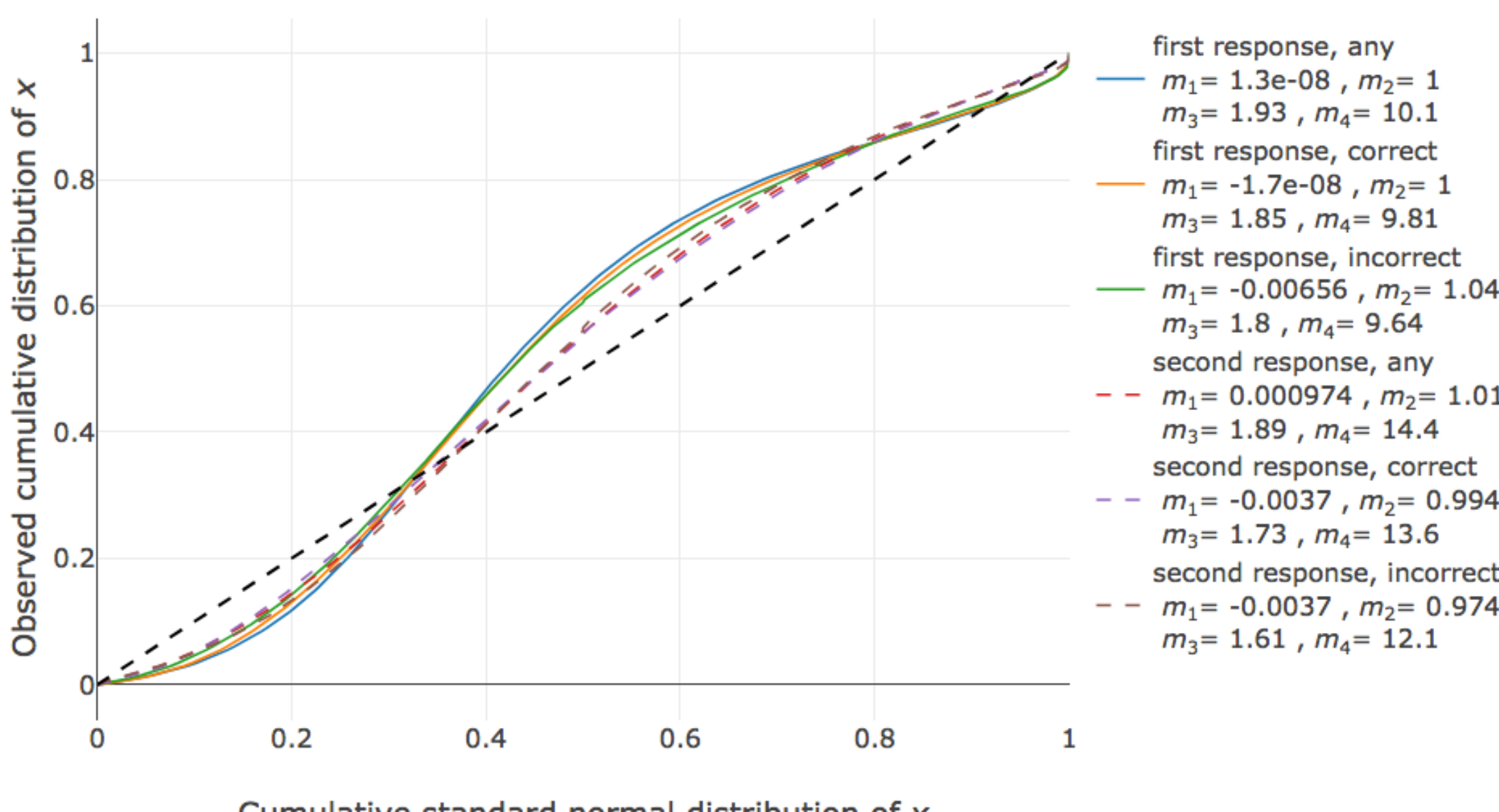

**Figure 1: Comparison of the observed cumulative distribution of the values $x_{qu}$ to the standard normal distribution $\Phi(x)$, predicted by Eq. 1. The identity line (shown in dashed black) represents the ideal agreement with the model. The listed distribution moments are calculated with respect to the ideal mean 0.**

The curves in Figure 1 appear to form two groups based on the submit number, whereas the submit correctness has a lesser effect. In essence, we can focus on the data coming from first and second submits of any correctness, and use the correctness-specific data to get the idea of the uncertainty size. The distribution of the first response times has a much smaller excess kurtosis and skewness than the distribution of the second response times, and almost perfect first and second moments. In all cases, the skewness and the excess kurtosis are positive (meaning that the sample distribution has heavier tails than the model predicts).

Selecting one row or column in the data matrices gives the distributions of $x_{qu}$ by question or by user. To quantify how frequently large deviations from Eq. 1 occur, we calculate the deviations of the first four moments of the distributions by question from the standard normal distribution, namely the quantities $d_k = (m_k)^{1/k} - \left(m_k^{(0)}\right)^{1/k}$ for $k = 1, 2, 3, 4$, where $m_k^{(0)} = (0,1,0,3)$ are the central moments of the standard normal distribution. Thus, $d_1$ is the mean, $d_2$ is the excess of standard deviation, $d_3$ is skewness, and $d_4$ is a measure similar to excess kurtosis. In Figure 2, we plot the percentile curves for these quantities.

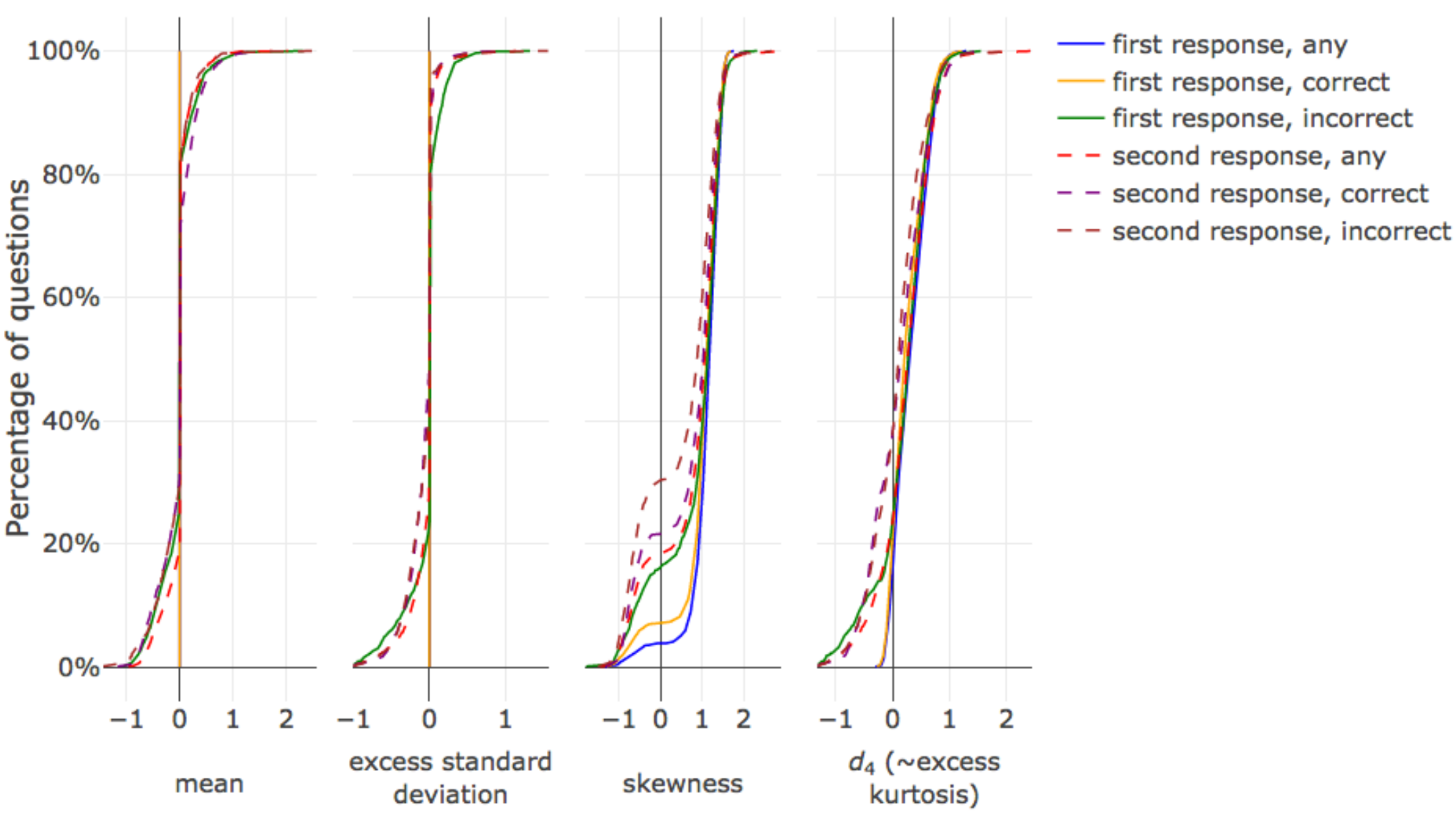

**Figure 2: Parameters of distributions observed for response time logarithms for each question. Ideal agreement with the model would mean all curves collapsing onto unit-step functions. In fact, the $d_1$ (mean) and $d_2$ (excess standard deviation) curves for the first submits of any correctness and for the first correct submits (blue and orange) are so close to the step function that the deviations are hard to notice in the image.**

Let us now investigate the outputs of the model: the user slowness $\zeta_u$ and the question parameters $\alpha_q$ and $\beta_q$. How do they depend on the response correctness and on the attempt number? Before aggregating across courses, it is instructive to examine one course as an example. In a given course, there is some overlap in users and questions in the data of first and second responses of different correctness, which allows comparing the parameters $\alpha_q$ and $\beta_q$ for the same question (or the slowness $\zeta_u$ for the same user) but obtained from different subsets of data. Using a STEM course with high degree of such an overlap in the data as a representative example, we plot its model parameters from first responses of different correctness, and from first and second responses (Figures 3 and 4). Here again, submit correctness is not a major factor: the points in Figure 3 cluster around the $y = x$ line and show substantial correlation, although the correlation of $\alpha_q$ always proves to be the lowest of the three (hence, whatever effect the correctness has, it is primarily on the degree of variability of the response times). Correlations between second correct and incorrect submits are lower, but otherwise the picture is similar. On the other hand, the difference between the models estimated on first and second submits is big (Figure 4). We expect that the typical time spent on the second submit is much shorter than on the first, so it is no surprise that the points for time intensity cluster well below the $x = y$ line. More surprisingly, the discrimination tends to increase on the second responses (less variability in the second response times).

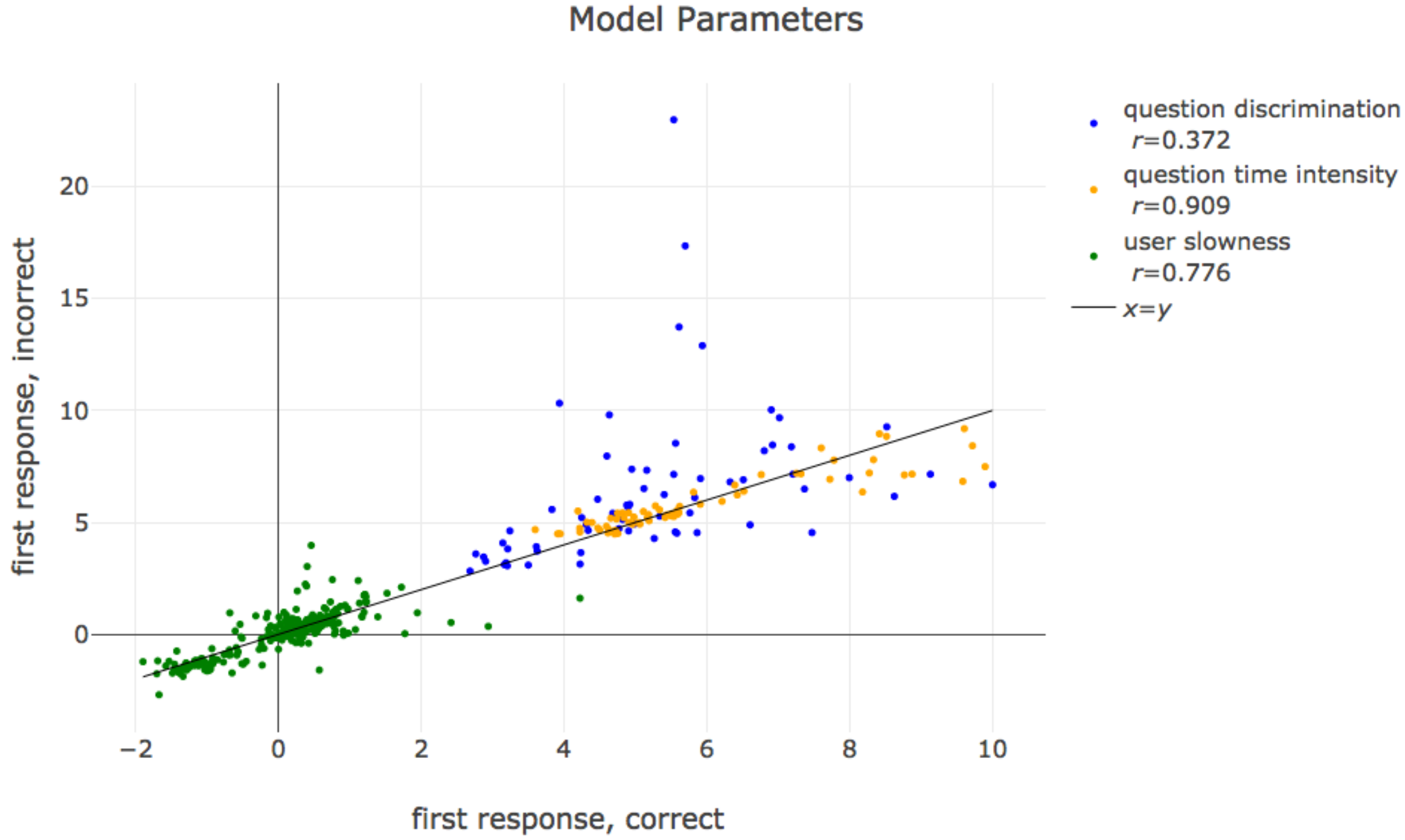

**Figure 3: A STEM course example. Model parameters obtained from first submits, correct vs. incorrect submits. Blue points are question discriminations $\alpha_q$, multiplied by 10 for better visibility. Yellow points are question time intensities $\beta_q$. Green points are user slownesses $\zeta_u$, multiplied by 0.5. The $r$ values are the correlations of values on the $x$ and $y$ axes.**

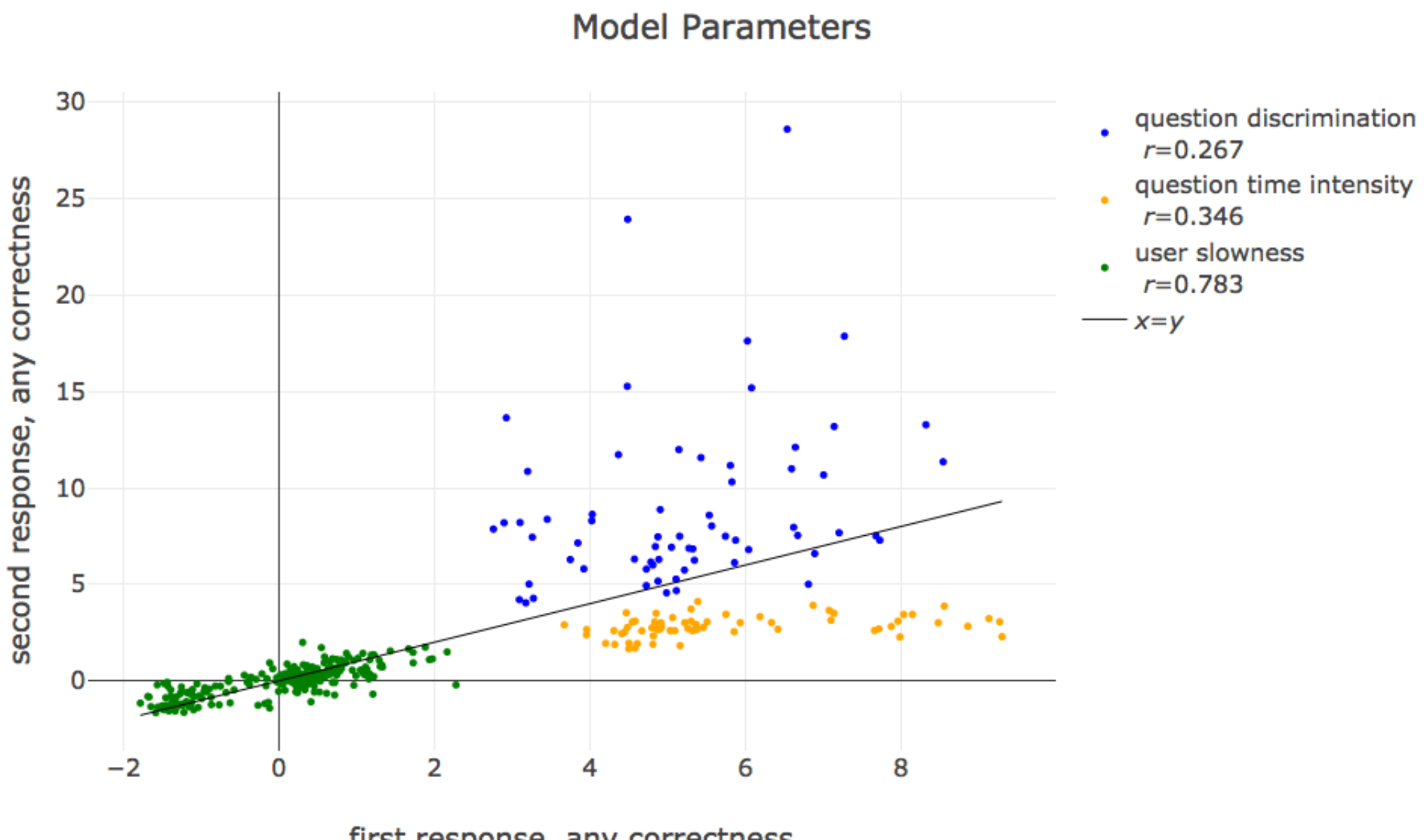

**Figure 4: A STEM course example. Model parameters obtained from first submits and second submits of any correctness. Blue points are question discriminations $\alpha_q$, multiplied by 10 for better visibility. Yellow points are question time intensities $\beta_q$. Green points are user slownesses $\zeta_u$, multiplied by 0.5. The $r$ values are the correlations of values on the $x$ and $y$ axes.**

While the differences between first and second submits may seem moderate in the plots, they translate into very sizeable time differences: in this course data, the median response time was 123 seconds on the first submit and 4 seconds on the second submit.

Figures 5 and 6 show the distribution densities for $\alpha_q$ and $\beta_q$ obtained from all converged data subsets from all courses. The densities are calculated by Gaussian-kernel smoothing. They reiterate the conclusions drawn from Figures 3 and 4: the time intensity tends to be smaller for the second submit (i.e., the second responses tend to be much quicker) but the discrimination is higher (although it has a broader distribution across questions). The distributions of time intensities on second submits are bimodal, with smaller peaks at $\beta = 2$ for STEM and $\beta = 3$ for non-STEM courses. Exponentiated, these correspond to typical response times of 7 and 20 seconds. These peaks are due to questions where users tend to make a small change in the answer and quickly resubmit (e.g., changing the sign in the numeric answer in a STEM course). The main distribution peaks lie near $\beta = 3.5$ ($\exp \beta \approx 33$ seconds) for all courses.[8]

Table 2 summarizes the correlations between model parameters, as obtained from different subsets of data. This is analogous to the correlation coefficients in Figures 3 and 4, but for all courses in the dataset. The observed pattern is the same. Slowness and time intensity are not very sensitive to response correctness (high correlation, especially on first submits), but the correlation between the data from first and second submits is lower.

**Table 2: Pearson Correlation Coefficient of Model Parameters Obtained from Different Data Subsets, All Courses**

| | User slowness $\zeta_u$ | Question time intensity $\beta_q$ | Question discrimination $\alpha_q$ |
|---|---|---|---|
| Correct vs. incorrect, first submits | $0.656 \pm 0.012$ | $0.698 \pm 0.014$ | $0.281 \pm 0.026$ |
| Correct vs. incorrect, second submits | $0.605 \pm 0.007$ | $0.869 \pm 0.008$ | $0.406 \pm 0.027$ |
| First vs. second submits, any correctness | $0.469 \pm 0.006$ | $0.186 \pm 0.018$ | $-0.139 \pm 0.019$ |

[8] We also repeated the data analysis using the page-load timestamp to calculate first response times, i.e., ignoring the fact that this extends the response times when multiple questions are served on the same page. In this case, the first response time intensity distribution also becomes bimodal, but for a different reason. The main peak is around $\beta = 8$ and the secondary peak was around $\beta = \ln(24 \cdot 3600) \approx 11.4$, due to users loading a page with multiple questions and working on some of them the next day.

**Note:** One standard error of the correlation coefficient is shown after "±."

The median time intensity of a question across all courses is 5.098 and 3.155 on the first and second responses, respectively (of any correctness). Exponentiated, these become the user-averaged typical response times of 164 seconds and 23 seconds, respectively. The median question discriminations on first and second responses (of any correctness) are 0.511 and 0.691. As a reminder, discrimination is the inverse standard deviation of the distribution of time logarithms on a given question. Exponentiating the inverses of these values, we find 7.08 on the first submits and 4.25 on the second. For instance, we can describe in rough terms (made precise by Eq. 1 and the medians of distributions in Figures 5 and 6) the situation for second submits as follows: after an unsuccessful first submit on a typical question, a typical user is expected to spend 23 seconds before submitting a second answer. The user variability is such that for most users the actual time lies in the range between $23/4.25 \approx 5$ seconds and $23 \cdot 4.25 \approx 98$ seconds. Obviously, this is a broad range, which after all is the reason that we choose time logarithm, rather than time itself, as the model variable.

**Figure 5: Distribution density of question discriminations $\alpha_q$. Responses of any correctness.**

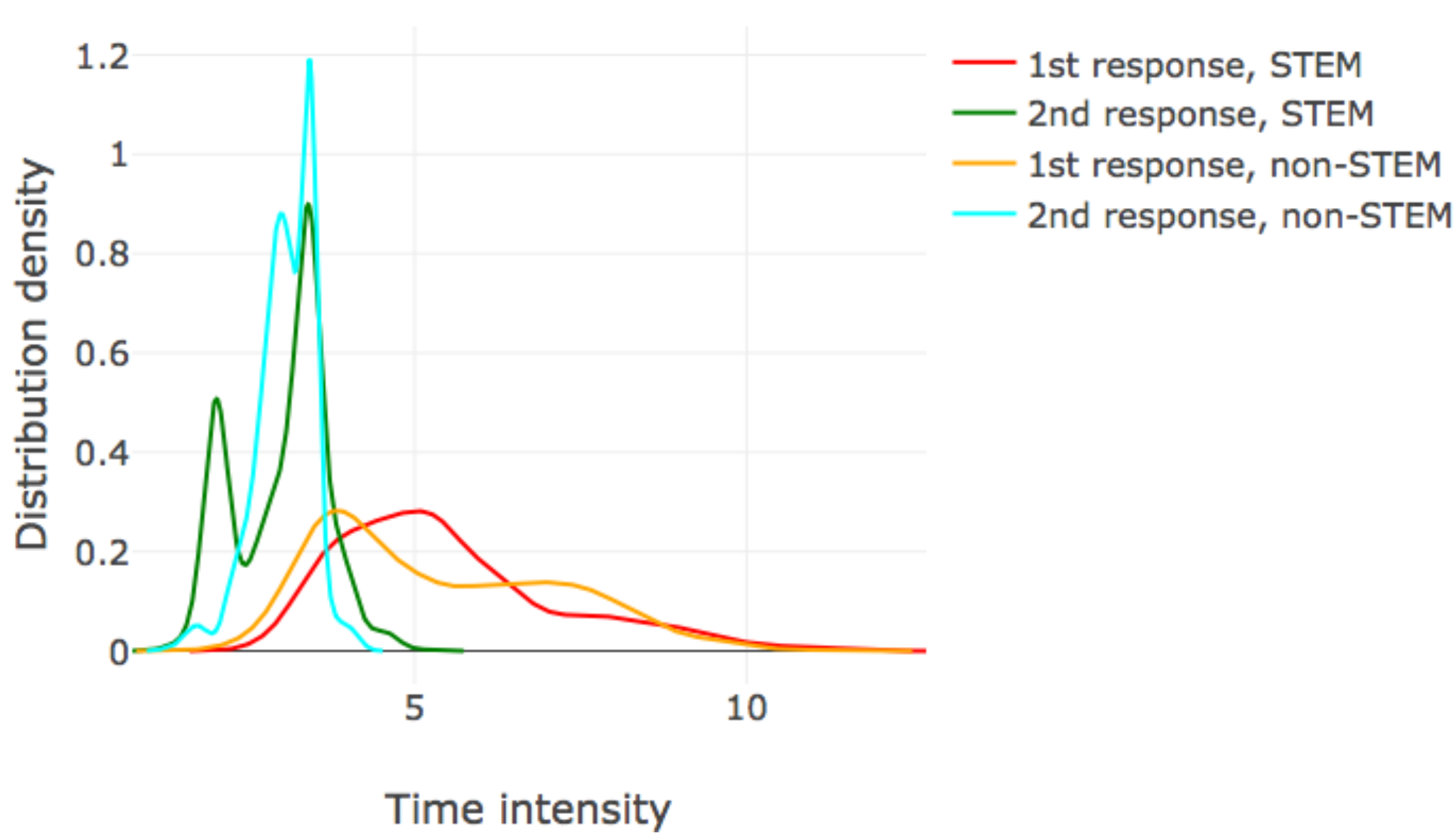

**Figure 6: Distribution density of question time intensity $\beta_q$. Responses of any correctness.**

Finally, it is interesting to analyze the relationship between $\alpha_q$ and $\beta_q$: what part of the observed increase in $\alpha_q$ is due specifically to the second attempt on the question and what part is a simple corollary of the lower $\beta_q$ (observable for quicker questions even on the same submit attempt)? Indeed, we find in our data (Figure 7) that on first responses $\beta_q$ and $1/\alpha_q$ are positively correlated ($\rho = 0.63$). In other words, if a question's response times are shorter on average, so is the multiplicative spread. A likely explanation is that quicker questions are not just scaled-down versions of slower ones. They are of a simpler nature, less open-ended or with fewer alternative solution paths, which decreases the variability in response times.

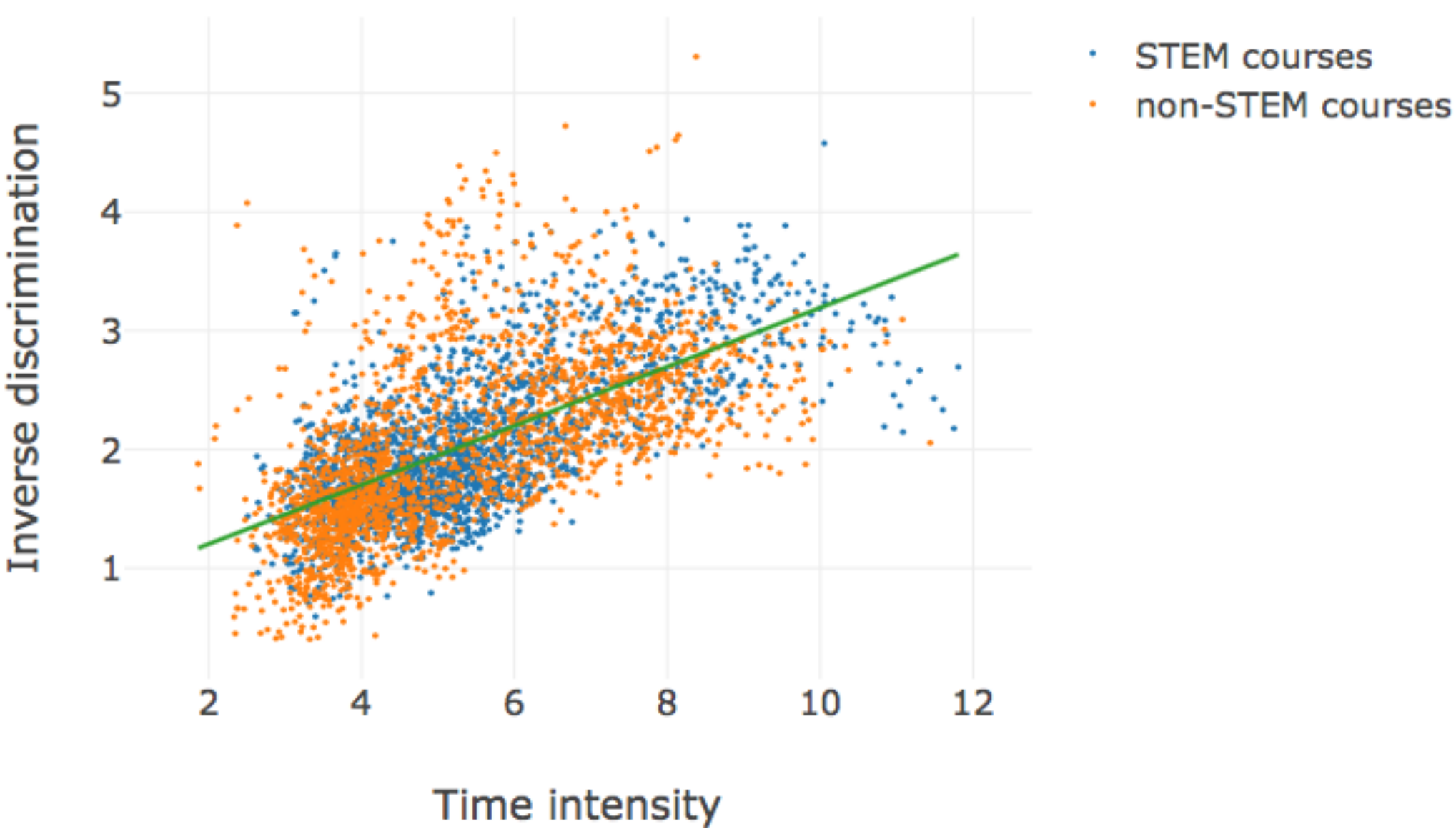

**Figure 7: A plot of $\beta_q$ vs. $1/\alpha_q$, for all courses. Data from first responses of any correctness.**

## 6. USER SLOWNESS AS A PREDICTOR OF COURSE OUTCOME

Once we obtain the slowness value for each course user, we want to investigate how this variable is connected to the user's success in the course. Our three main measures of user success are the final grade (a numeric score on 0-to-1 scale), course completion (defined as achieving a final grade not lower than the passing grade, which is set for each course by the course instructors), and earning the course certificate. We trained a set of mixed-effect linear regression models, one with each of these as the outcome variable (for an introduction to such models, see Goldstein, 2011). For the final grade (models "Grade 1" and "Grade 2"), we fit linear models by restricted maximum likelihood; the p-values are obtained with the t-test in Satterthwaite's (1946) approximation. Completion and certification are binary outcomes fitted to generalized linear models by maximum likelihood in Laplace approximation, with the logistic link function.[9] The models allowed for random effect of courses, thus accounting for the diversity of courses in our data set. Our key independent variables of interest are $\zeta^{(1)}$ and $\zeta^{(2)}$: user slowness, obtained from either the first or the second submits of any correctness. To control for the user's proficiency in the course, we included the fraction of correct responses on the first submit ("Correctness") and the self-declared level of education ("Education"). The level of education was an ordinal variable formed as follows: 0 indicates no education (<1% of users); 1 indicates elementary school (<1%); 2 indicates junior high school (2%); 3 indicates high school (13%); 4 indicates associate degree (4%), 5 indicates bachelor's (31%); 6 indicates master's (38%); 7 indicates PhD (7%). The results are summarized in Table 3.

---

[9] Coefficients of logistic regression (used in the models for completion and certification) can be interpreted in terms of the outcome odds (defined as $p/(1-p)$ where $p$ is the outcome probability). For instance, coefficient 0.102 for $\zeta^{(1)}$ in the model "Completion 1" means that, all other variables being equal, increasing slowness by 1 is associated with multiplying the odds of completion by $\exp(0.102) \approx 1.107$, i.e., an increase of about 11%.

**Table 3: Summary of Course Outcome Models: All Models Allow for Random Effect of Diverse Courses in Addition to the Predictors Listed**

| Model | Predictor | St. dev. | Coef. ± Std. err. | $p$ |
|---|---|---|---|---|
| Completion 1: Completion $\sim\zeta^{(1)}$ + Correctness + Education | $\zeta^{(1)}$ | 1.162 | 0.102 ± 0.011 | $<10^{-10}$ |
| | Correctness | 0.366 | 4.098 ± 0.189 | $<10^{-10}$ |
| | Education | 1.289 | −0.002 ± 0.010 | 0.865 |
| Certification 1: Certification $\sim\zeta^{(1)}$ + Correctness + Education | $\zeta^{(1)}$ | 1.162 | 0.133 ± 0.012 | $<10^{-10}$ |
| | Correctness | 0.366 | 3.065 ± 0.206 | $<10^{-10}$ |
| | Education | 1.289 | −0.055 ± 0.011 | $8{\cdot}10^{-7}$ |
| Grade 1: Final grade $\sim\zeta^{(1)}$ + Correctness + Education | $\zeta^{(1)}$ | 1.162 | 0.003 ± 0.002 | 0.056 |
| | Correctness | 0.366 | 0.633 ± 0.022 | $<10^{-10}$ |
| | Education | 1.289 | 0.001 ± 0.001 | 0.384 |
| Completion 2: Completion $\sim\zeta^{(1)} + \zeta^{(2)}$ + Correctness + Education | $\zeta^{(1)}$ | 1.083 | −0.113 ± 0.018 | $5{\cdot}10^{-10}$ |
| | $\zeta^{(2)}$ | 1.046 | 0.515 ± 0.021 | $<10^{-10}$ |
| | Correctness | 0.261 | 3.495 ± 0.276 | $<10^{-10}$ |
| | Education | 1.243 | 0.008 ± 0.014 | 0.567 |
| Certification 2: Certification $\sim\zeta^{(1)} + \zeta^{(2)}$ + Correctness + Education | $\zeta^{(1)}$ | 1.083 | 0.005 ± 0.018 | 0.801 |
| | $\zeta^{(2)}$ | 1.046 | 0.334 ± 0.019 | $<10^{-10}$ |
| | Correctness | 0.261 | 2.156 ± 0.289 | $<10^{-10}$ |
| | Education | 1.243 | −0.032 ± 0.015 | 0.029 |
| Grade 2: Final grade $\sim\zeta^{(1)} + \zeta^{(2)}$ + Correctness + Education | $\zeta^{(1)}$ | 1.083 | −0.022 ± 0.002 | $<10^{-10}$ |
| | $\zeta^{(2)}$ | 1.046 | 0.068 ± 0.002 | $<10^{-10}$ |
| | Correctness | 0.261 | 0.545 ± 0.031 | $<10^{-10}$ |
| | Education | 1.243 | −0.001 ± 0.002 | 0.490 |

The effect of the education level turns out to be either small or statistically insignificant. The most important predictor in all three models is, unsurprisingly, the correctness fraction: people who do well on assessments are more likely to finish the course with a good grade. But after that is taken into account, slowness has a sizeable and positive effect, meaning that users who take more time to submit an answer are more likely to complete the course with a higher grade and to get a certificate. The coefficient on $\zeta^{(2)}$ is particularly large (in models "Completion 2," "Certification 2," "Grade 2"), while is $\zeta^{(1)}$ small or even negative (most pronounced in "Completion 2"). This negative effect of $\zeta^{(1)}$ does not contradict the positive effects in "Completion 1," "Certification 1," and "Grade 1," since the models with $\zeta^{(2)}$ are trained on the data events when the second attempt was submitted after an incorrect first attempt, which is a biased subset of the full data on which the models with only $\zeta^{(1)}$ are trained. It can be interpreted as follows: if a user submitted a wrong answer on a first attempt and then made a second attempt at a question, it is more likely for course-completers that this happened because the first wrong answer was given relatively quickly (perhaps, rashly?) and the second answer was provided slowly. We have already excluded users who did not explore courses or interacted with few questions, so the results are not dominated by users who are not committed to learning and who simply click through.

## 7. USER SLOWNESS IN RELATION TO OTHER USER VARIABLES

We explore the correlation of user slowness with other user variables available to us, again by fitting mixed-effect linear models with random effect of courses, but this time using slowness as the outcome variable. Some of the predictor variables are the descriptors of the level of engagement in the course: number of videos viewed in the course ("Videos"), number of "play video" clicks ("Play clicks"), and the number of forum posts ("Posts"). We also add two demographic variables: the self-reported level of education ("Education," same as we used before) and age at the start date of the course ("Age"). The mean age of users in our data was 38 years, with a standard deviation of 15 years.

Because different courses contain different numbers of videos and encourage forum use to different degrees, we normalized the variables "Videos," "Play clicks," "Posts" to unit mean value across users in each course prior to fitting the model. Furthermore, to make the coefficients of the model easier to interpret as correlations, we rescale all the variables in the model to unit variance within each course. The results are summarized in Table 4.

**Table 4: Summary of User-Slowness Models**

| Model | Predictor | Coef. ± Std. err. | $p$ |
|---|---|---|---|
| Slowness 1: $\zeta^{(1)} \sim$ Education + Age + Videos + Play clicks + Posts | Education | $-0.015 \pm 0.009$ | 0.093 |
| | Age | $0.112 \pm 0.009$ | $<10^{-10}$ |
| | Videos | $0.044 \pm 0.009$ | $8 \cdot 10^{-7}$ |
| | Play clicks | $0.129 \pm 0.007$ | $<10^{-10}$ |
| | Posts | $0.086 \pm 0.008$ | $<10^{-10}$ |
| Slowness 2: $\zeta^{(2)} \sim$ Education + Age + Videos + Play clicks + Posts | Education | $-0.011 \pm 0.011$ | 0.345 |
| | Age | $0.195 \pm 0.011$ | $<10^{-10}$ |
| | Videos | $-0.042 \pm 0.011$ | $2 \cdot 10^{-4}$ |
| | Play clicks | $0.074 \pm 0.009$ | $<10^{-10}$ |
| | Posts | $0.083 \pm 0.011$ | $<10^{-10}$ |

**Note:** All models allow for random effect of diverse courses, in addition to the predictors listed.

The level of education turns out to have no significant effect on slowness. But we see that slowness has a mild positive correlation with the user's age and, in case of slowness on the first attempt, with the number of "play video" clicks, which implies that users who take longer on assessment questions also have a tendency to pause-and-play videos more.

The overall conclusion is that greater slowness is associated with higher achievement (measured by final course grades) and engagement (measured by completion, certification, and watching videos). Moreover, there is a small positive correlation of slowness with user age but not with level of education.

## 8. DISCUSSION AND FUTURE WORK

The described log-normal model of response times can be used to estimate the characteristics of both the learners (slowness) and the assessment questions (discrimination and time intensity). In contrast to earlier findings in similar models as applied to test data, we find that in self-paced online courses from our dataset that higher user slowness is linked to higher achievement levels: students who take their time tend to do better. Since slowness (or some other metric reflecting learner response times) can be tracked by the course instructor in the course analytics in addition to other performance metrics with the goal of flagging struggling learners, our result suggests that faster and low-performing learners are of greater concern than slower ones. We also find that user slowness and time intensity of questions can be estimated from response times regardless of response correctness, since the separate estimates from correct and incorrect responses turn out to be strongly correlated.

It is worth noting that the response times, which we model in this work, are distinct from the concept of time on task, which is much harder to estimate, or even define precisely (Cetintas, Si, Xin, & Hord, 2010; Grabe & Sigler, 2002). It is not necessary to assume that the response time is spent entirely in the "on task" mode. In fact, we assume (although this is not a required assumption either) that distractions are one of the sources of variability in response times.

The question characteristics, extracted from the model, may be useful in course design. Questions are commonly transferred, with no or minimal alterations, from one version of a course to the next. Examining the time intensity and discrimination of each question in previous iterations of the course can alert instructors to questions that are outliers either in time intensity or in discrimination, which may suggest that certain questions are too hard or too easy.

The discovered relation between user slowness and success brings up several questions, which we hope to investigate in future, that have to do with the practice of online learning. The most straightforward one is to track learner slowness and design an intervention for those who move fast without achieving high assessment scores, suggesting that they should take more time on questions. A/B testing with such an intervention would clarify the causal direction of the relationship between slowness and success. Another interesting research question is to clarify the possible reasons, and their relative importance, for users taking an unusually long or short time on a question. Learners can be asked to select from a list of possible explanations in a survey. A third future direction of study is the application of our model to course design. The time-intensity (and, possibly, discrimination as well) of assessment items offer the course team insight into the way learners interact with the assessment. For instance, items that are outliers in terms of time-intensity would be flagged. So, if a question intended by the authors as a quick check turns out to have a large time-intensity, that question could be reviewed or even replaced.

## ACKNOWLEDGMENTS

We are grateful to the Office of the Vice Provost for Advances in Learning at Harvard University for thoughtful leadership and support.

## DECLARATION OF CONFLICTING INTEREST

The authors declared no potential conflicts of interest with respect to the research, authorship, and/or publication of this article.

## FUNDING

The authors declared no financial support for the research, authorship, and/or publication of this article.